\documentclass[conference]{IEEEtran}
\IEEEoverridecommandlockouts

\usepackage{cite}
\usepackage{amsmath,amssymb,amsfonts}
\usepackage{algorithmic}
\usepackage{graphicx}
\usepackage{textcomp}
\usepackage{xcolor}
\usepackage{xurl}
\usepackage{microtype}
\usepackage[hidelinks]{hyperref}

\def\BibTeX{{\rm B\kern-.05em{\sc i\kern-.025em b}\kern-.08em
    T\kern-.1667em\lower.7ex\hbox{E}\kern-.125emX}}
\begin{document}

\title{Nepali Passport Question Answering: A Low-Resource Dataset for Public Service Applications\\}

\author{
\IEEEauthorblockN{1\textsuperscript{st} Funghang Limbu Begha}
\IEEEauthorblockA{
\textit{ILPRL}\\
\textit{Department of Computer Science \& Engineering}\\
Kathmandu University, Dhulikhel, Nepal\\
funghang.limbu.3@gmail.com
}
\and
\IEEEauthorblockN{2\textsuperscript{nd} Praveen Acharya}
\IEEEauthorblockA{
\textit{School of Computing}\\
\textit{Dublin City University}\\
Dublin, Ireland\\
acharyaprvn@gmail.com
}
\and
\IEEEauthorblockN{3\textsuperscript{rd} Bal Krishna Bal\thanks{Corresponding author: bal@ku.edu.np}}
\IEEEauthorblockA{
\textit{ILPRL}\\
\textit{Department of Computer Science \& Engineering}\\
Kathmandu University, Dhulikhel, Nepal\\
bal@ku.edu.np
}
}

\maketitle
\begin{abstract}
Nepali, a low-resource language, faces significant challenges in building an effective information retrieval system due to the unavailability of annotated data and computational linguistic resources. In this study, we attempt to address this gap by preparing a pair-structured  Nepali Question-Answer dataset. We focus on Frequently Asked Questions (FAQs) for passport-related services, building a data set for training and evaluation of embedding-based IR models. In our study, we have fine-tuned transformer-based embedding models for semantic similarity in question-answer retrieval. The fine-tuned models were compared with the lexical baseline BM25. In addition, we implement a hybrid retrieval approach, integrating fine-tuned models with BM25, and evaluate the performance of the hybrid retrieval. Our results show that the fine-tuned embedding-based models outperform BM25, whereas multilingual E5 embedding-based retrieval models achieve the highest retrieval performance among all evaluated models. 
\end{abstract}


\section{Introduction}
Information Retrieval (IR) systems have made remarkable progress in the development of transformer-based language models such as BERT and its variants~\cite{devlin2019bert,reimers2019sentence}. These models have enabled embedding-based retrieval methods that capture deep semantic relationships between queries and documents~\cite{karpukhin2020dense}. However, such progress has been largely prevalent in high-resource languages, making low-resource languages like Nepali highly disadvantaged. 

Previous studies on Nepali Natural Language Processing (NLP) have revealed that the language still lacks large-scale annotated datasets, standardized benchmarks, and domain-specific linguistic resources required to support advanced NLP applications~\cite{shahi2022natural}. Shahi and Sitaula (2021)~\cite{shahi2022natural} comprehensively reviewed the state of Nepali NLP and emphasized that most research has focused on higher-level tasks like classification and sentiment analysis, while fundamental resources such as gold-standard datasets and text representations remain underdeveloped and underexplored. This limitation continues to hinder progress in subsequent areas, such as information retrieval, question answering, and other semantic tasks essential for digital government services and citizen information systems.

To address these gaps, this study introduces a framework for Nepali question-answering retrieval focused on frequently asked questions (FAQs) related to passport services. We developed a pair-structured Nepali question-answering dataset to facilitate training and evaluation. Multiple multilingual transformer-based embedding models were fine-tuned for semantic question-answer retrieval using the Sentence Transformer framework and compared against the BM25 lexical retrieval baseline. 

The main contributions of this work are as follows.
\begin{itemize}
  \item Developed a Nepali question-answering dataset focused on passport services, enabling domain-specific retrieval research.
  \item Fine-tuned embedding-based transformer models (BERT, RoBERTa, and E5) using Multiple Negative Ranking Loss (MNRL)~\cite{henderson2017efficient} for semantic question answer retrieval and evaluated their performance.
\end{itemize}

\section{Related Work}
\label{sec:related_work}
Traditional information retrieval methods such as TF–IDF~\cite{salton1988term} and BM25~\cite{robertson2009probabilistic} have long served as the foundation for lexical retrieval by matching exact terms between queries and documents. Although BM25~\cite{robertson2009probabilistic} remains a strong baseline for many retrieval tasks due to its efficiency, it cannot capture the semantic relationship between queries and documents.

Recent advances in neural embedding-based retrieval have focused on dense retrieval models that capture semantic similarity through contextual embeddings. Sentence-BERT (SBERT)~\cite{reimers2019sentence} introduced a bi-encoder architecture to generate fixed-size sentence embeddings optimized for semantic similarity tasks. The multilingual SBERT~\cite{reimers2020making} enables cross-lingual retrieval and transfer learning, which are useful in low-resource environment. Similarly, Dense Passage Retrieval (DPR)~\cite{karpukhin2020dense} demonstrated how dual encoder architectures can outperform traditional lexical models in open-domain question answering by dense representation of queries and documents.

In the context of Nepali, several sentence embedding models have been proposed for semantic similarity tasks. This includes Yunika/sentence transformer-nepali~\cite{yunika_sentence_transformer}, universalml/Nepali\_Embedding\_Model~\cite{universalml_nepali_embedding_model}, jangedoo/all-MiniLM-L6-v2-nepali~\cite{jangedoo_minilm_nepali}, Syubraj/sentence\_similarity\_nepali~\cite{syubraj_sentence_similarity_nepali}, and the multilingual variants of intfloat/e5 (small, base and large)~\cite{wang2024multilingual}.

Some of these models have been fine-tuned specifically for Nepali semantic similarity tasks, while the multilingual intfloat/e5 variants were pre-trained for general semantic similarity. However, none of these models have been systematically evaluated for information retrieval in domain-specific datasets, such as Nepali question answering for passport services. More recently, Pudasaini et al.~\cite{pudasaini2025nepaligpt} introduced NepaliGPT, a generative language model for Nepali, along with a general large Devanagari Corpus and a Nepali question-answer dataset. Their work illustrates the increasing availability of Nepali-specific resources for NLP and highlights potential applications in both generative and retrieval-based systems.

In addition,~\cite{bajracharya2025extractive} introduced a Nepali extractive Question Answering System to help overcome the scarcity of Nepali Question Answering (QA) datasets. They contributed three main resources including a Nepali and Hindi translation of SQuAD 1.1~\cite{rajpurkar2016squad}, a Nepali translation of XQuAD~\cite{artetxe2019cross} for evaluation, and a newly compiled Nepali QA dataset derived from Belebele’s MCQ data~\cite{bandarkar2023belebele}. Their work focuses on extractive question answering based on span, demonstrating that fine-tuning multilingual models with translation-invariant tokens significantly improves performance on Nepali QA benchmarks. Although their dataset emphasizes extractive comprehension, our work is different in that we construct a domain-specific native Nepali question-answer retrieval dataset designed for semantic question–answer retrieval rather than span extraction.

Recent work has also investigated Nepali question-answering systems using transformer models. Thapa et al.~\cite{thapa2024nepali} developed a Nepali QA system by fine-tuning multilingual BERT (mBERT) and monolingual BERT (NepBERTa) models on a Nepali QA dataset derived from SQuAD. Their study highlights the effectiveness of transformer-based models for low-resource languages, showing that mBERT outperforms NepBERTa in terms of F1 and BLEU scores. This work underscores the potential of pre-trained multilingual and monolingual models in addressing Nepali-specific QA and retrieval tasks, and motivates further exploration of dense retrieval approaches for domain-specific corpora.

In recent years, several benchmark datasets have been proposed for the retrieval and development of FAQs. De et al.~\cite{de2021mfaq} introduced a large-scale multilingual FAQs dataset covering multiple domains, while COUGH~\cite{zhang2020cough} provided a specialized dataset for COVID-19 FAQs, allowing evaluation of retrieval systems in domain-specific contexts. More recently, WebFAQ~\cite{dinzinger2025webfaq} presented a comprehensive multilingual collection of natural Q\&A pairs designed for dense retrieval, highlighting the importance of multilingual FAQ resources. Although these datasets primarily target high-resource languages, their design and objectives motivate the creation of Nepali-specific QA datasets.

For modeling approaches, recent research has explored both dense and hybrid retrieval strategies for FAQ and domain-specific question-answer systems. MFBE~\cite{banerjee2023mfbe} leveraged multi-field information (question, answer, metadata) for efficient dense retrieval in industrial FAQs. Seo et al.~\cite{seo2022dense} proposed a hybrid retriever that combines sparse BM25 and dense (transformer-based) components, balancing lexical precision with semantic generalization. Similarly, Domain-specific Question Answering with Hybrid Search~\cite{sultania2024domain} demonstrated that hybrid models that integrate BM25 and dense embeddings improve retrieval quality in specialized domains. Rayo et al.~\cite{rayo2025hybrid} confirmed that the combination of lexical and semantic representations improves performance for complex domain-regulated datasets.

From a retrieval strategy perspective, prior work has also explored semantic matching between queries, FAQ questions, and answers. Sakata et al.~\cite{sakata2019faq} proposed a BERT-based system that measures both how closely a query matches a FAQ question and how it is relevant to the associated answer, showing improved FAQ retrieval performance compared to purely lexical methods.

In the Nepali language domain, Poudel et al.~\cite{poudel2023retrieval} developed a pregnancy-related chatbot that compared retrieval-based methods using multilingual BERT and DistilBERT with generative transformers. Their findings indicated that BERT-based pre-trained models perform well on non-stemmed data and the stemmed transformer-based generative model also performs effectively on the pregnancy-related dataset, though evaluation on domain-specific retrieval tasks is limited. A RAG-based Nepali legal question-answering system was developed~\cite{bhusal2025enhanced}, focusing on questions related to the Public Service Commission (PSC) examination held in Nepal. They used bigram matching, the OpenAI text-embedding-3-large embedding model, and a combined approach for retrieval, with GPT-4 for answer generation. Their results showed that the bigram matching retriever outperformed the semantic text-embedding-3-large model, while the combined approach outperformed both bigram matching and the semantic text-embedding-3-large model. Furthermore, GPT-4 performed best, achieving a BERT F1 score of 0.69. Wagle et al.~\cite{wagle2026retrieval} also developed a RAG-based question-answering system focusing on the Nepali legal domain. The system used BM25 for lexical retrieval and multilingual E5 variants and LaBSE for dense retrieval, with GPT-o3 for answer generation. Their results showed that BM25-based retrieval, particularly BM25-Chunk, outperformed dense retrieval models. However, BM25-Docs was used in the RAG pipeline due to its lower retrieval latency. BM25-Docs achieved the best overall generation performance, with a 92\% successful answer generation rate. In addition to domain-specific Nepali QA and RAG systems, Acharya and Bal~\cite{acharya2026towards} highlighted the lack of standardized evaluation resources for Nepali information retrieval and proposed a benchmark evaluation framework covering monolingual and cross-lingual retrieval. The proposed framework aims to systematically evaluate sparse, dense, and hybrid retrieval methods using query-document relevance judgments.

\section{DATASET AND PREPROCESSING}
\subsection{Data Collection}
\begin{figure}[ht!]
     \centering
     \includegraphics[scale=0.42]{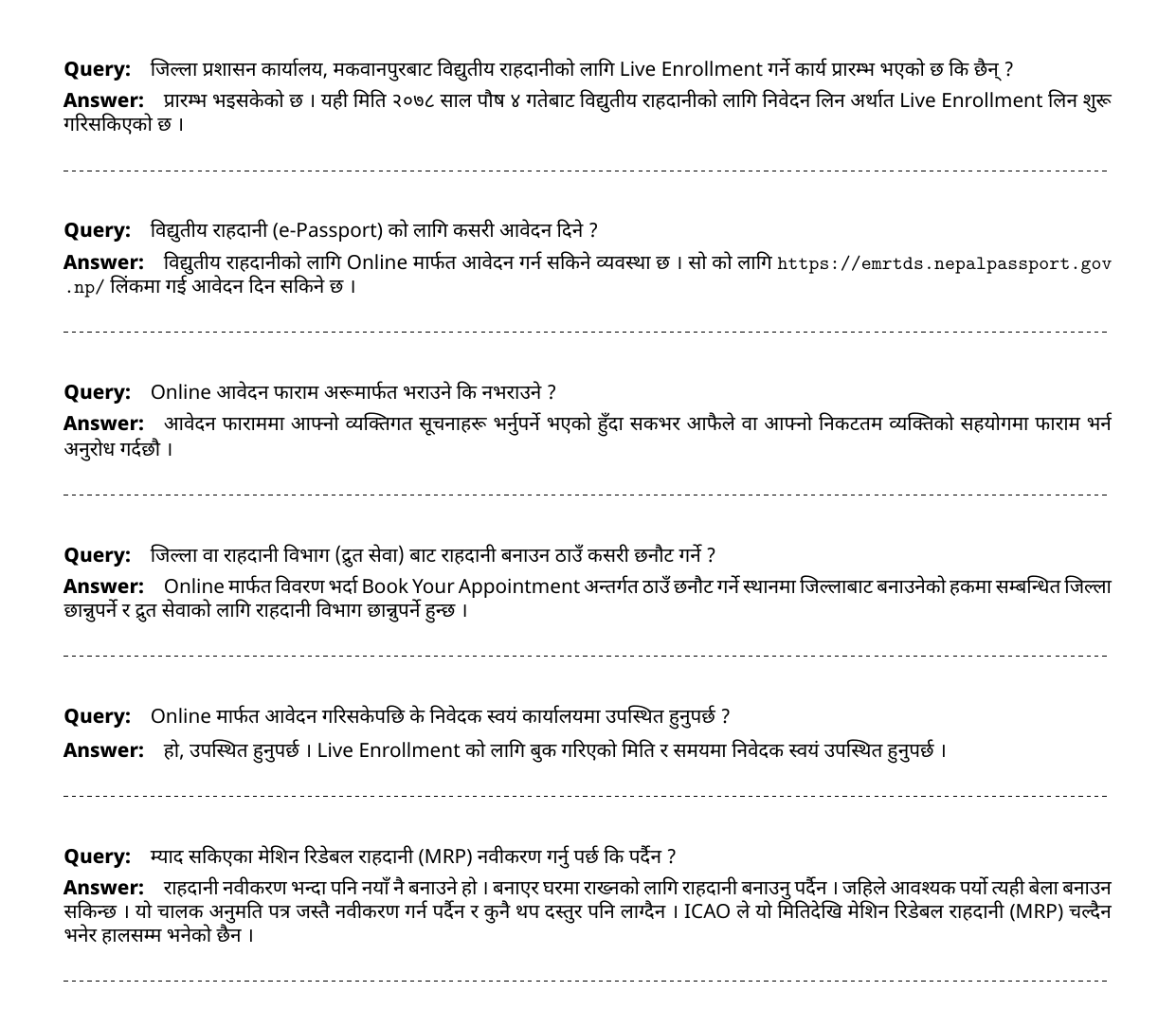}
     \caption{Sample of FAQs from Our Dataset}
     \label{fig:Extracted_data}
\end{figure}

To fill the gap in domain-specific Nepali Information Retrieval, we focus on data collection and compiling comprehensive datasets of question-answer pairs related to passports and services provided by the Ministry of Foreign Affairs in Nepal. The FAQs were scraped from official government websites, PDF documents, and public resources to ensure the accuracy of the domain. We develop automated web scraping tools and scripts to crawl the specified websites and extract relevant Passport FAQs. The scraping process targeted FAQ sections, help pages, and other areas of websites where passport-related information was published. A sample of the dataset scraped from the relevant sources is shown in Figure~\ref{fig:Extracted_data}. 

\subsection{Data Pre-processing}
\label{sec:data_preprocessing}
For data pre-processing, we performed several steps to ensure the quality and consistency of our Nepali FAQ dataset. Initially, we removed HTML tags, URLs, and special characters from all text entries. Next, we standardized Unicode to normalize Nepali characters, ensuring consistent encoding for NLP tasks. Common spelling errors and inconsistent formatting were corrected. Duplicate entries were removed to reduce redundancy. We follow both a manual approach and a cosine-similarity-based approach to check the duplicate queries and answers. First, we compute the similarity score across all queries and answers, select the query pairs and answer pairs with a higher score, and manually check the duplicate entries. Our dataset is a collection of question-answer pairs, where each entry contains a query representing a user question and a corresponding positive field representing the correct answer. To further enhance the dataset and address low-resource constraints, we applied data augmentation using GPT-4, generating additional variations of FAQ pairs while preserving semantic meaning. All queries and answers in the train/val split and the answers of the test queries were augmented. 

\subsection{Data Analysis}

\begin{table}[htbp]
\centering
\caption{Statistics of the Nepali Passport Question-Answering Dataset}
\label{tab:dataset_statistics}
\small
\setlength{\tabcolsep}{6pt} 
\renewcommand{\arraystretch}{1.15} 
\begin{tabular}{|l|c|c|c|c|}
\hline

\multicolumn{5}{|c|}{\textbf{Original QA Pairs}} \\ \hline
\textbf{Features} & \textbf{Train} & \textbf{Validation} & \textbf{Test} & \textbf{Total/Avg} \\ \hline
Query & 384 & 82 & 82 & 548 \\ \hline
Answer & 384 & 82 & 82 & 548 \\ \hline
Tokens per Query & 16.94 & 17.17 & 16.17 & 16.86 \\ \hline
Tokens per Answer & 47.39 & 49.49 & 54.65 & 48.79 \\ \hline

\multicolumn{5}{|c|}{\textbf{Augmented QA Pairs}} \\ \hline
\textbf{Features} & \textbf{Train} & \textbf{Validation} & \textbf{Test} & \textbf{Total} \\ \hline
Query & 3840 & 820 & 82 & 4742 \\ \hline
Answer & 3840 & 820 & 820 & 5480 \\ \hline
Tokens per Query & 16.65 & 17.08 & 16.32 & 16.66 \\ \hline
Tokens per Answer & 42.20 & 45.07 & 47.64 & 43.44 \\ \hline
\end{tabular}

\end{table}

\begin{table}[htbp]
\centering
\caption{Nepali Passport Question-Answering Retrieval Test Set Statistics}
\label{tab:ir_testset}
\begin{tabular}{lcccc}
\hline
\textbf{Query} & \textbf{Relevancy} & \textbf{Corpus} & \textbf{Avg D/Q} & \textbf{Avg Word Length} \\
\hline
82 & Binary & 37,013 & 10 & 10.9 \\
\hline
\end{tabular}
\end{table}

\begin{figure}[htbp]
     \centering
     \includegraphics[width=\columnwidth]{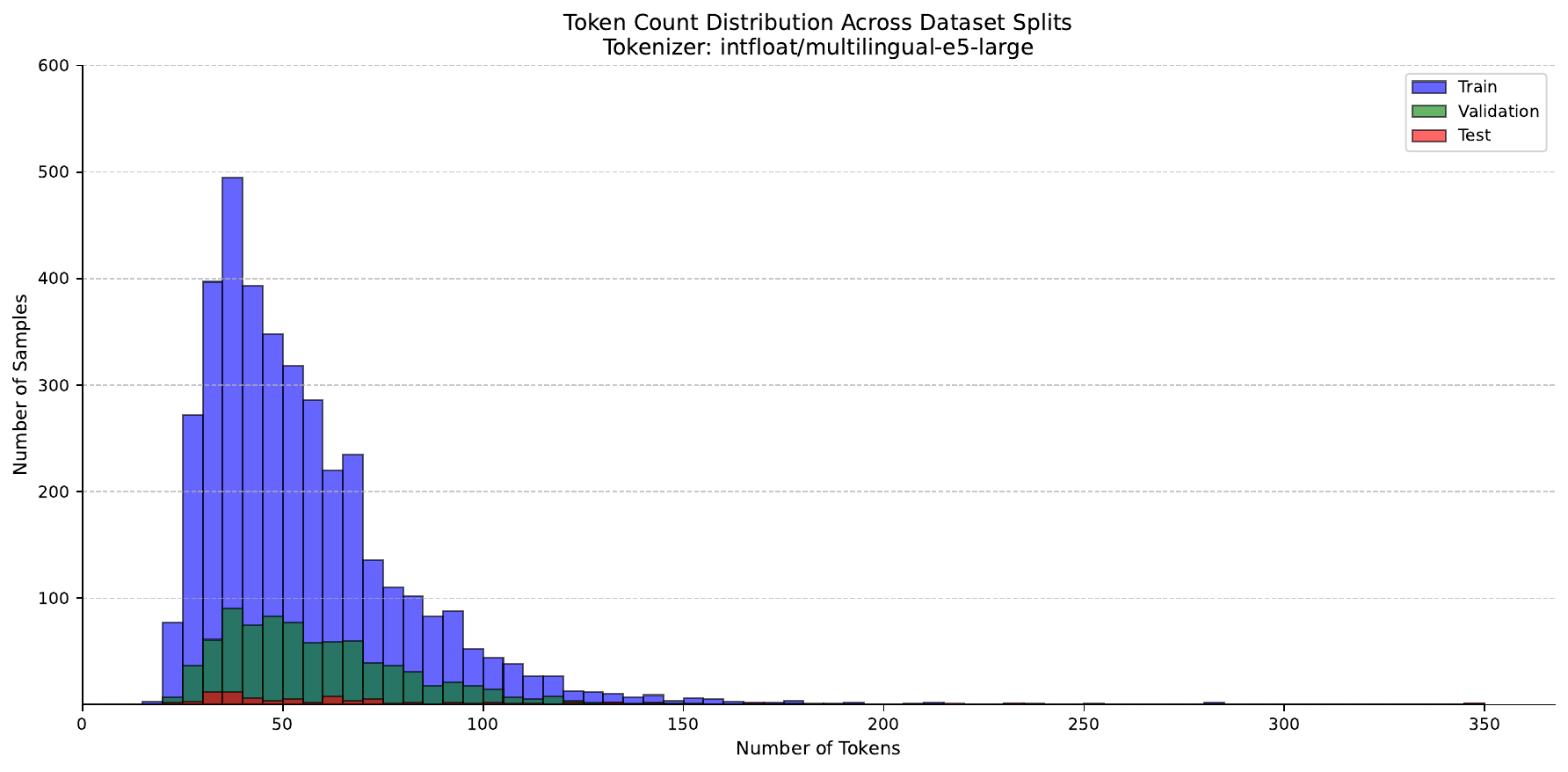}
     \caption{Token count distribution across the Nepali question-answer pair dataset for training, validation, and test splits using the intfloat/multilingual-e5-large tokenizer. The histogram shows the number of tokens per pair, calculated by summing the tokens of the query and its corresponding positive entry, highlighting the overall sequence length patterns in the dataset.}
     \label{fig:Data_distribution}
\end{figure}

The dataset used in this study consists of 548 unique FAQs related to Nepali passport services. These original FAQs were carefully divided into training, validation, and test sets. Table~\ref{tab:dataset_statistics} presents the statistics of the augmented question–answer dataset, including the number of queries, answers, and the average number of tokens per query and per answer in the train, validation, and test sets. The augmentation increased the dataset size by roughly tenfold while preserving consistent token length distributions across splits.

For evaluation, we built a test set based on the BEIR dataset~\cite{thakur2021beir}. The test set includes 82 queries, each linked to a set of 10 relevant documents, allowing binary relevance judgments. To simulate a realistic retrieval environment, we combined 820 relevant documents (test answers) with 36,193 irrelevant documents obtained from the publicly available raygx/Nepali-Extended-Text-Corpus~\cite{raygx_Nepali_Extended_Text_Corpus}. The combination produced a total corpus of 37,013 documents.

Table~\ref{tab:ir_testset} summarizes the statistics of our test set, including the number of queries, relevance type, corpus size, average number of documents per query, and average word length per document. Our test corpus represents a challenging evaluation scenario with a high ratio of irrelevant to relevant document, reflecting real-world document retrieval conditions. The dataset was further analyzed to understand its token level distribution. Figure~\ref{fig:Data_distribution} presents a bar graph showing the number of tokens versus the frequency in training, validation, and the test set.

\section{Methodology}
\begin{figure*}[htbp]
    \centering
    \includegraphics[width=\textwidth]{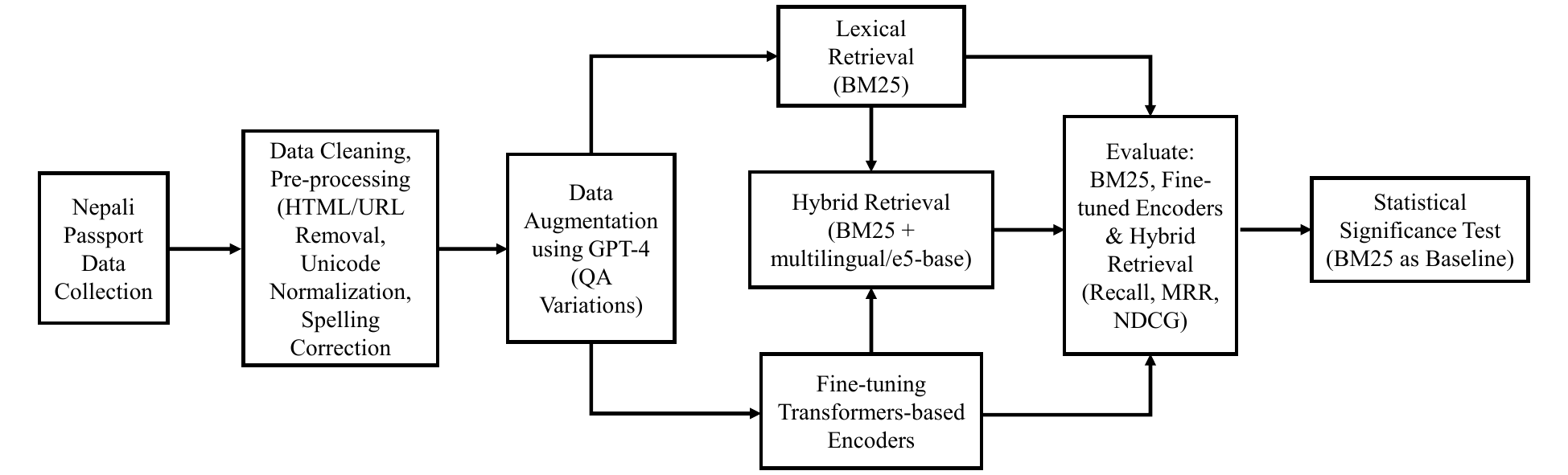}
    \caption{Workflow of the proposed information retrieval framework, which evaluates lexical (BM25), fine-tuned embedding-based model, and hybrid (BM25 + intfloat/multilingual-e5-base) retrieval models, followed by statistical significance testing against the BM25 baseline.}
    \label{fig:workflow}
\end{figure*}
The workflow of our Nepali question answering retrieval framework is shown in Figure~\ref{fig:workflow}. Our framework integrates both lexical and embedding-based retrieval approaches to handle lexical and semantic similarity in low-resource environment. Our framework begins with data pre-processing and augmentation of question-answer pairs, followed by fine-tuning multilingual embedding models on the prepared dataset. For evaluation, we conducted retrieval evaluation using BM25, fine-tuned embedding models, and hybrid approaches. Finally, statistical significance tests (Paired t-test~\cite{student1908probable} and Wilcoxon Signed-Rank Test~\cite{wilcoxon1992individual}) are performed against the baseline BM25 to assess the significance of observed improvements.

\subsection{Model selection}
For lexical retrieval, we chose the lexical BM25 ranking function, and for semantic retrieval embedding-based models were chosen for their ability to capture semantic similarity. Our selection includes pre-trained multilingual models as well as models fine-tuned for Nepali semantic similarity as discussed in Section~\ref{sec:related_work}. This includes SBERT-based embedding models (BERT and RoBERTa) and variants of the E5 embedding model optimized for cross-lingual semantic embeddings.

We excluded models that lacked pretraining and fine-tuning on the Nepali dataset. The models chosen above allow for comparison between lexical, dense embedding, and hybrid retrieval approaches in a domain-specific Nepali dataset.

\subsection{Training Setup}
\label{sec:training_setup}
The selected models were fine-tuned in our dataset with a consistent training configuration to ensure fair comparison. we have set an identical optimization strategy and regularization parameters for each model and the batch size and learning rates were optimized through grid search. Table~\ref{tab:common_training_config} shows the common hyperparameters used in training.

\begin{table}[htbp]
\centering
\caption{Common Training Configuration Across All Models}
\label{tab:common_training_config}
\small
\setlength{\tabcolsep}{8pt} 
\renewcommand{\arraystretch}{1.1} 
\begin{tabular}{|l|l|}
\hline
\textbf{Parameter} & \textbf{Value} \\ \hline
Evaluation Strategy & Steps \\ \hline
Evaluation Steps & 100 \\ \hline
Warmup ratio & 0.01 \\ \hline
Fp16 & True \\ \hline
\end{tabular}
\end{table}

An Adamw optimizer with a linear learning rate and Multiple Negative Ranking Loss (MNRL)~\cite{henderson2017efficient} was used for training. A pair-structured QA dataset is used for training where negatives are implicitly sampled within the batch, eliminating the need for explicit triplet or negative mining. An early stopping mechanism was applied to prevent overfitting with a patience of 5. The models used for fine-tuning and their respective optimized hyperparameters obtained from the grid search are shown in Table~\ref{tab:fine_tune_config}.

\section{Experimental Setup}
\subsection{Environment Configuration}
\subsubsection{Hardware Setup}
The models were trained and evaluated using high performance GPUs to accelerate embedding computations and model fine-tuning. Specifically, we used a T4 GPU provided by Google Colab and an L4 GPU provided by Lightning AI. These GPUs enabled efficient training of multiple embedding based models.

\subsubsection{Software Setup}
Training and evaluation were performed using Python 3.12.12 in Google Colab. The key libraries and frameworks used include Transformers, PyTorch, BeautifulSoup, Pandas, NumPy, and Matplotlib.

\subsection{Experimental Procedure}
In this section, we have described the steps followed to fine-tune and evaluate the embedding-based retrieval models discussed in Table~\ref{tab:fine_tune_config}. During data preparation, our Nepali question-answer dataset had already been split as described in Section~\ref{sec:data_preprocessing}. These models were fine-tuned with the training configuration discussed in Section~\ref{sec:training_setup}.
\begin{table}[htbp]
\centering
\caption{Fine-tuning Configurations for All Embedding Models Used in Our Study}
\label{tab:fine_tune_config}
\scriptsize 
\setlength{\tabcolsep}{7pt} 
\renewcommand{\arraystretch}{1.4} 
\begin{tabular}{|p{4.5cm}|c|c|} 
\hline
\textbf{Model} & \textbf{Batch Size} & \textbf{Learning Rate} \\ \hline
Yunika/sentence-transformer-nepali~\cite{yunika_sentence_transformer} & 8 & $1\times10^{-5}$ \\ \hline
Syubraj/sentence\_similarity\_nepali~\cite{syubraj_sentence_similarity_nepali} & 8 & $2\times10^{-5}$ \\ \hline
universalml/Nepali\_Embedding\_Model~\cite{universalml_nepali_embedding_model} & 8 & $1\times10^{-5}$ \\ \hline
jangedoo/all-MiniLM-L6-v2-nepali~\cite{jangedoo_minilm_nepali} & 8 & $1\times10^{-5}$ \\ \hline
intfloat/e5-small (multilingual)~\cite{wang2024multilingual} & 8 & $4\times10^{-5}$ \\ \hline
intfloat/e5-base (multilingual)~\cite{wang2024multilingual} & 8 & $3\times10^{-5}$ \\ \hline
intfloat/e5-large (multilingual)~\cite{wang2024multilingual} & 8 & $1\times10^{-5}$ \\ \hline
\end{tabular}
\end{table}

During the training phase, the evaluation checkpoints were recorded every 100 steps, selecting the checkpoint with the lowest validation loss for the final test. The query and document pairs were represented as dense embeddings, and the cosine similarity between the query and the answer was calculated to rank the candidate document. We implemented batch processing during both training and evaluation to handle multiple query-document pairs simultaneously.

\subsection{Evaluation Setup}
We evaluate the retrieval model using the InformationRetrievalEvaluator class of the Sentence-Transformers library~\cite{reimers2019sentence}. The evaluator ranks documents based on similarity scores and compares the ranked results with ground truth relevance labels. The evaluation metrics are Recall\text{@}k, Mean Reciprocal Rank (MRR\text{@}k), and Normalized Discounted Cumulative Gain (NDCG\text{@}k).

\section{RESULT AND DISCUSSION}
\subsection{Retrieval Performance}
Table~\ref{tab:results_significance} presents the results in Recall\text{@}k, Mean Reciprocal Rank MRR\text{@}k and Normalized Discounted Cumulative Gain NDCG\text{@}k for $k = 5$ and $k = 10$. From the result, the lexical baseline BM25 achieved Recall\text{@}5 of 0.3537 and Recall\text{@}10 of 0.4793 showing moderate effectiveness in retrieving relevant answers. In general, embedding-based models significantly outperform the BM25 in all metrics.

\begin{table*}[!ht]
\centering
\caption{Evaluation Results of Retrieval Models on Nepali Passport Question-Answering Dataset. Superscripts indicate statistical significance against BM25 baseline: $\alpha: p<0.05$, $\beta: p<0.01$. Values without superscripts are not statistically significant.}
\label{tab:results_significance}
\small
\setlength{\tabcolsep}{4pt}
\renewcommand{\arraystretch}{1}
\resizebox{\textwidth}{!}{%
\begin{tabular}{lcccccc}
\hline
\textbf{Model} & \textbf{Recall@5} & \textbf{Recall@10} & \textbf{NDCG@5} & \textbf{NDCG@10} & \textbf{MRR@5} & \textbf{MRR@10} \\ 
\hline
BM25 & \multicolumn{1}{c}{0.3537} & \multicolumn{1}{c}{0.4793} & \multicolumn{1}{c}{0.7393} & \multicolumn{1}{c}{0.5693} & \multicolumn{1}{c}{0.8689} & \multicolumn{1}{c}{0.8757} \\ 
\hline
Syubraj/sentence\_similarity\_nepali & \multicolumn{1}{c}{0.3683} & 0.5951$^\beta$ & \multicolumn{1}{c}{0.7599} & 0.6546$^\beta$ & \multicolumn{1}{c}{0.8598} & \multicolumn{1}{c}{0.8598} \\
Yunika/sentence-transformer-nepali & 0.4671$^\beta$ & 0.7988$^\beta$ & 0.9464$^\beta$ & 0.8493$^\beta$ & 0.9858$^\alpha$ & 0.9858$^\alpha$ \\
jangedoo/all-MiniLM-L6-v2-nepali & \multicolumn{1}{c}{0.3598} & 0.6037$^\beta$ & \multicolumn{1}{c}{0.7361} & 0.6510$^\alpha$ & \multicolumn{1}{c}{0.8327} & \multicolumn{1}{c}{0.8388} \\
intfloat/e5-small & 0.4695$^\beta$ & 0.8232$^\beta$ & 0.9472$^\beta$ & 0.8663$^\beta$ & 0.9760$^\beta$ & 0.9760$^\beta$ \\
intfloat/e5-base & 0.4841$^\beta$ & 0.8573$^\beta$ & \textbf{0.9756}$^\beta$ & 0.8973$^\beta$ & \textbf{1.0000}$^\beta$ & \textbf{1.0000}$^\beta$ \\
intfloat/e5-large & \textbf{0.4878}$^\beta$ & \textbf{0.8902}$^\beta$ & 0.9754$^\beta$ & \textbf{0.9188}$^\beta$ & 0.9817$^\beta$ & 0.9817$^\beta$ \\
universalml/Nepali\_Embedding\_Model & 0.4780$^\beta$ & 0.8598$^\beta$ & 0.9653$^\beta$ & 0.8965$^\beta$ & 0.9939$^\beta$ & 0.9939$^\beta$ \\ 
\hline
BM25 + intfloat/e5-base & 0.4805$^\beta$ & 0.8317$^\beta$ & 0.9701$^\beta$ & 0.8785$^\beta$ & \textbf{1.0000}$^\beta$ & \textbf{1.0000}$^\beta$ \\ 
\hline
\end{tabular}%
}
\end{table*}

Embedding-based retrieval models~\cite{syubraj_sentence_similarity_nepali, jangedoo_minilm_nepali} show small improvements over BM25, achieving Recall\text{@}10 of 0.5951 and 0.6037, respectively. Another semantic embedding model~\cite{yunika_sentence_transformer} shows a significant performance boost, achieving Recall\text{@}10 of 0.7988 and NDCG\text{@}10 of 0.8493, reflecting the effectiveness of Nepali-specific fine-tuning for semantic similarity. 

Among the multilingual variants of E5~\cite{wang2024multilingual}, a trend of increased performance was observed with increasing model size. The E5-small achieved a Recall\text{@}10 of 0.8232 and NDCG\text{@}10 of 0.8663. The improved base variant of E5 reaches Recall\text{@}10 of 0.8573 and NDCG\text{@}10 of 0.8973, achieving perfect MRR\text{@}5 and MRR\text{@}10 (1.0000), indicating that the correct answer was ranked first. Although the E5-large slightly outperformed the E5-base in Recall\text{@}10 of 0.8902 and NDCG\text{@}10 of 0.9188, its MRR\text{@}10 of 0.9817 was marginally lower than the E5-base. The model introduced in~\cite{universalml_nepali_embedding_model} also performed competitively with Recall\text{@}10 of 0.8598 and NDCG\text{@}10 of 0.8965, close to the base variant of E5~\cite{wang2024multilingual}, reflecting the benefits of language specific training.

The hybrid retrieval model (BM25 + intfloat/multilingual-e5-base) further improves performance by integrating lexical and semantic matching. It achieves a Recall\text{@}10 of 0.8317 and NDCG\text{@}10 of 0.8785, which is slightly lower than the E5-base~\cite{wang2024multilingual} alone in Recall\text{@}10. Furthermore, the hybrid approach achieves a perfect MRR\text{@}10 of 1.0000, indicating that for certain queries, this approach ensures that the top-ranked answer is correct.

\subsection{Discussion}  
In general, our findings demonstrate that embedding-based and hybrid retrieval methods are more effective than the BM25 ranking function, particularly for low-resource languages such as Nepali. SBERT-based models outperformed BM25 in individual comparisons. In addition, the multilingual E5 embedding variants are also suitable for semantic question-answer retrieval in Nepali language. Among the multilingual variants of E5, E5-base outperformed all others, but the E5-small and E5-large performed better than other SBERT-based models. The hybrid approach model (BM25 + intfloat/multilingual-e5-base) achieves the best balance between lexical precision and semantic relevance, establishing it as the reliable retrieval framework for the Nepali passport FAQ dataset.

\section{Conclusion and Future Work}
In this study, we performed research on a domain-specific question-answer retrieval task in Nepali language. Multiple multilingual embedding models, including~\cite{syubraj_sentence_similarity_nepali, yunika_sentence_transformer, jangedoo_minilm_nepali, universalml_nepali_embedding_model} and  variants of~\cite{wang2024multilingual}, were systematically compared with the lexical baseline of BM25. Our experimental results demonstrated that embedding-based models outperform the lexical baseline in all evaluation metrics. The E5-base~\cite{wang2024multilingual} and the universalml/Nepali\_Embedding\_Model~\cite{universalml_nepali_embedding_model} achieved near-perfect retrieval performance, highlighting the effectiveness of multilingual and fine-tuned embeddings for Nepali NLP tasks.

Furthermore, the hybrid approach (BM25 + intfloat/multi\allowbreak lingual-e5-base) achieved improved performance over BM25, combining the lexical precision of BM25 with the semantic strength of the multilingual E5-base. This integration resulted in improved retrieval performance and ranking metrics, confirming that hybrid retrieval effectively leverages contextual similarities. These findings indicate that hybrid retrieval models are particularly well-suited for languages such as Nepali, where purely lexical models alone may be insufficient.

This study also emphasized the importance of preparing a dataset, including the construction of a pair-structured Nepali FAQ dataset with relevant and irrelevant examples. This work establishes a framework for developing and evaluating retrieval systems in domain-specific and low-resource contexts.

In conclusion, this study demonstrates that embedding-based retrieval and hybrid retrieval approaches significantly outperform traditional lexical methods for Nepali question-answering retrieval. The proposed dataset and evaluation framework provide a strong foundation for the development of low-resource domain-specific retrieval systems.

In addition, research can extend this work by exploring Retrieval-Augmented Generation (RAG) frameworks\cite{lewis2020retrieval} that combine lexical and semantic retrieval with generative models such as GPT\cite{achiam2023gpt} or mT5\cite{xue2020mt5}. Although this study focused only on retrieval models, generative models can produce contextually rich and natural answers directly from queries. RAG frameworks further enhance this capability by retrieving relevant documents and generating a contextually rich and natural response based on the retrieved context. Such models can utilize the retrieved context as input to generative decoders, generating more natural responses for open-domain queries. Further studies could also examine the adaptability of the proposed framework to other Nepali domains, including healthcare, education, or government services, to assess transferability between domains.

\section*{Acknowledgment}

 Special thanks to the Department of Computer Science and Engineering, Kathmandu University, and the Information and Language Processing Research Lab (ILPRL) for providing computational resources and a supportive research environment.

\bibliographystyle{IEEEtran}
\bibliography{references}  

@article{reimers2019sentence,
  title={Sentence-bert: Sentence embeddings using siamese bert-networks},
  author={Reimers, Nils and Gurevych, Iryna},
  journal={arXiv preprint arXiv:1908.10084},
  year={2019}
}

@article{robertson2009probabilistic,
  title={The probabilistic relevance framework: BM25 and beyond},
  author={Robertson, Stephen and Zaragoza, Hugo and others},
  journal={Foundations and Trends{\textregistered} in Information Retrieval},
  volume={3},
  number={4},
  pages={333--389},
  year={2009},
  publisher={Now Publishers, Inc.}
}

@article{xue2020mt5,
  title={mT5: A massively multilingual pre-trained text-to-text transformer},
  author={Xue, Linting and Constant, Noah and Roberts, Adam and Kale, Mihir and Al-Rfou, Rami and Siddhant, Aditya and Barua, Aditya and Raffel, Colin},
  journal={arXiv preprint arXiv:2010.11934},
  year={2020}
}

@inproceedings{devlin2019bert,
  title={Bert: Pre-training of deep bidirectional transformers for language understanding},
  author={Devlin, Jacob and Chang, Ming-Wei and Lee, Kenton and Toutanova, Kristina},
  booktitle={Proceedings of the 2019 conference of the North American chapter of the association for computational linguistics: human language technologies, volume 1 (long and short papers)},
  pages={4171--4186},
  year={2019}
}

@inproceedings{karpukhin2020dense,
  title={Dense Passage Retrieval for Open-Domain Question Answering.},
  author={Karpukhin, Vladimir and Oguz, Barlas and Min, Sewon and Lewis, Patrick SH and Wu, Ledell and Edunov, Sergey and Chen, Danqi and Yih, Wen-tau},
  booktitle={EMNLP (1)},
  pages={6769--6781},
  year={2020}
}

@article{wang2024multilingual,
  title={Multilingual e5 text embeddings: A technical report},
  author={Wang, Liang and Yang, Nan and Huang, Xiaolong and Yang, Linjun and Majumder, Rangan and Wei, Furu},
  journal={arXiv preprint arXiv:2402.05672},
  year={2024}
}

@misc{yunika_sentence_transformer,
  author = {Yunika Bajracharya},
  title = {Yunika/sentence-transformer-nepali},
  year = {2024},
  url = {https://huggingface.co/Yunika/sentence-transformer-nepali},
  note = {Hugging Face model}
}

@misc{jangedoo_minilm_nepali,
  title={jangedoo/all-MiniLM-L6-v2-nepali},
  author={Sanjaya Subedi},
  url ={https://huggingface.co/jangedoo/all-MiniLM-L6-v2-nepali/tree/main},
  year={2024}
}

@misc{syubraj_sentence_similarity_nepali, 
  title={Syubraj/sentence\_similarity\_nepali}, 
  author={Yubraj Sigdel}, 
  url ={https://huggingface.co/syubraj/sentence_similarity_nepali}, 
  year={2023} 
}

@misc{raygx_Nepali_Extended_Text_Corpus,
  title={raygx/Nepali-Extended-Text-Corpus},
  author={Regan Maharjan},
  year={2023},
  url={https://huggingface.co/datasets/raygx/Nepali-Extended-Text-Corpus},
}

@misc{universalml_nepali_embedding_model,
  title={universalml/Nepali\_Embedding\_Model},
  author={Singh Prince},
  url ={https://huggingface.co/universalml/Nepali_Embedding_Model},
  year={2024}
}

@article{reimers2020making,
  title={Making monolingual sentence embeddings multilingual using knowledge distillation},
  author={Reimers, Nils and Gurevych, Iryna},
  journal={arXiv preprint arXiv:2004.09813},
  year={2020}
}

@article{salton1988term,
  title={Term-weighting approaches in automatic text retrieval},
  author={Salton, Gerard and Buckley, Christopher},
  journal={Information processing \& management},
  volume={24},
  number={5},
  pages={513--523},
  year={1988},
  publisher={Elsevier}
}

@article{poudel2023retrieval,
  title={Retrieval and Generative Approaches for a Pregnancy Chatbot in Nepali with Stemmed and Non-Stemmed Data: A Comparative Study},
  author={Poudel, Sujan and Ghimire, Nabin and Subedi, Bipesh and Singh, Saugat},
  journal={arXiv preprint arXiv:2311.06898},
  year={2023}
}

@inproceedings{sakata2019faq,
  title={FAQ retrieval using query-question similarity and BERT-based query-answer relevance},
  author={Sakata, Wataru and Shibata, Tomohide and Tanaka, Ribeka and Kurohashi, Sadao},
  booktitle={Proceedings of the 42nd international ACM SIGIR conference on research and development in information retrieval},
  pages={1113--1116},
  year={2019}
}

@article{de2021mfaq,
  title={MFAQ: a multilingual FAQ dataset},
  author={De Bruyn, Maxime and Lotfi, Ehsan and Buhmann, Jeska and Daelemans, Walter},
  journal={arXiv preprint arXiv:2109.12870},
  year={2021}
}

@article{zhang2020cough,
  title={COUGH: A challenge dataset and models for COVID-19 FAQ retrieval},
  author={Zhang, Xinliang Frederick and Sun, Heming and Yue, Xiang and Lin, Simon and Sun, Huan},
  journal={arXiv preprint arXiv:2010.12800},
  year={2020}
}

@inproceedings{dinzinger2025webfaq,
  title={WebFAQ: A Multilingual Collection of Natural Q\&A Datasets for Dense Retrieval},
  author={Dinzinger, Michael and Caspari, Laura and Ghosh Dastidar, Kanishka and Mitrovi{\'c}, Jelena and Granitzer, Michael},
  booktitle={Proceedings of the 48th International ACM SIGIR Conference on Research and Development in Information Retrieval},
  pages={3802--3811},
  year={2025}
}

@inproceedings{banerjee2023mfbe,
  title={MFBE: Leveraging Multi-field Information of FAQs for Efficient Dense Retrieval},
  author={Banerjee, Debopriyo and Jain, Mausam and Kulkarni, Ashish},
  booktitle={Pacific-Asia Conference on Knowledge Discovery and Data Mining},
  pages={112--124},
  year={2023},
  organization={Springer}
}

@article{seo2022dense,
  title={Dense-to-question and sparse-to-answer: Hybrid retriever system for industrial frequently asked questions},
  author={Seo, Jaehyung and Lee, Taemin and Moon, Hyeonseok and Park, Chanjun and Eo, Sugyeong and Aiyanyo, Imatitikua D and Park, Kinam and So, Aram and Ahn, Sungmin and Park, Jeongbae},
  journal={Mathematics},
  volume={10},
  number={8},
  pages={1335},
  year={2022},
  publisher={MDPI}
}

@article{sultania2024domain,
  title={Domain-specific Question Answering with Hybrid Search},
  author={Sultania, Dewang and Lu, Zhaoyu and Naik, Twisha and Dernoncourt, Franck and Yoon, David Seunghyun and Sharma, Sanat and Bui, Trung and Gupta, Ashok and Vatsa, Tushar and Suresha, Suhas and others},
  journal={arXiv preprint arXiv:2412.03736},
  year={2024}
}

@article{shahi2022natural,
  title={Natural language processing for Nepali text: A review},
  author={Shahi, Tej Bahadur and Sitaula, Chiranjibi},
  journal={Artificial Intelligence Review},
  volume={55},
  number={4},
  pages={3401--3429},
  year={2022},
  publisher={Springer}
}

@article{thapa2024nepali,
  title={Nepali Question Answering System from Multilingual BERT Model and Monolingual BERT Model},
  author={Thapa, Upanshu and Timilsina, Suresh and Tiwari, Hom Nath and Upadhyay, Mukunda},
  year={2024}
}

@article{pudasaini2025nepaligpt,
  title={NepaliGPT: A Generative Language Model for the Nepali Language},
  author={Pudasaini, Shushanta and Shakya, Aman and Shrestha, Siddhartha and Bhatta, Sahil and Thapa, Sunil and Palikhe, Sushmita},
  journal={arXiv preprint arXiv:2506.16399},
  year={2025}
}

@article{bajracharya2025extractive,
  title={Extractive Nepali Question Answering System},
  author={Bajracharya, Yunika and Shrestha, Suban and Bastola, Saurav and Satyal, Sanjivan},
  journal={KEC Journal of Science and Engineering},
  volume={9},
  number={1},
  pages={95--102},
  year={2025}
}

@article{student1908probable,
  title={The probable error of a mean},
  author={Student},
  journal={Biometrika},
  pages={1--25},
  year={1908},
  publisher={JSTOR}
}

@incollection{wilcoxon1992individual,
  title={Individual comparisons by ranking methods},
  author={Wilcoxon, Frank},
  booktitle={Breakthroughs in statistics: Methodology and distribution},
  pages={196--202},
  year={1992},
  publisher={Springer}
}

@article{henderson2017efficient,
  title={Efficient natural language response suggestion for smart reply},
  author={Henderson, Matthew and Al-Rfou, Rami and Strope, Brian and Sung, Yun-Hsuan and Luk{\'a}cs, L{\'a}szl{\'o} and Guo, Ruiqi and Kumar, Sanjiv and Miklos, Balint and Kurzweil, Ray},
  journal={arXiv preprint arXiv:1705.00652},
  year={2017}
}

@article{thakur2021beir,
  title={Beir: A heterogenous benchmark for zero-shot evaluation of information retrieval models},
  author={Thakur, Nandan and Reimers, Nils and R{\"u}ckl{\'e}, Andreas and Srivastava, Abhishek and Gurevych, Iryna},
  journal={arXiv preprint arXiv:2104.08663},
  year={2021}
}

@inproceedings{rajpurkar2016squad,
  title={Squad: 100,000+ questions for machine comprehension of text},
  author={Rajpurkar, Pranav and Zhang, Jian and Lopyrev, Konstantin and Liang, Percy},
  booktitle={Proceedings of the 2016 conference on empirical methods in natural language processing},
  pages={2383--2392},
  year={2016}
}

@article{bandarkar2023belebele,
  title={The belebele benchmark: a parallel reading comprehension dataset in 122 language variants},
  author={Bandarkar, Lucas and Liang, Davis and Muller, Benjamin and Artetxe, Mikel and Shukla, Satya Narayan and Husa, Donald and Goyal, Naman and Krishnan, Abhinandan and Zettlemoyer, Luke and Khabsa, Madian},
  journal={arXiv preprint arXiv:2308.16884},
  year={2023}
}

@article{artetxe2019cross,
  title={On the cross-lingual transferability of monolingual representations},
  author={Artetxe, Mikel and Ruder, Sebastian and Yogatama, Dani},
  journal={arXiv preprint arXiv:1910.11856},
  year={2019}
}

@article{rayo2025hybrid,
  title={A Hybrid Approach to Information Retrieval and Answer Generation for Regulatory Texts},
  author={Rayo, Jhon and de La Rosa, Ra{\'u}l and Garrido, Mario},
  journal={arXiv preprint arXiv:2502.16767},
  year={2025}
}

@article{lewis2020retrieval,
  title={Retrieval-augmented generation for knowledge-intensive nlp tasks},
  author={Lewis, Patrick and Perez, Ethan and Piktus, Aleksandra and Petroni, Fabio and Karpukhin, Vladimir and Goyal, Naman and K{\"u}ttler, Heinrich and Lewis, Mike and Yih, Wen-tau and Rockt{\"a}schel, Tim and others},
  journal={Advances in neural information processing systems},
  volume={33},
  pages={9459--9474},
  year={2020}
}

@article{achiam2023gpt,
  title={Gpt-4 technical report},
  author={Achiam, Josh and Adler, Steven and Agarwal, Sandhini and Ahmad, Lama and Akkaya, Ilge and Aleman, Florencia Leoni and Almeida, Diogo and Altenschmidt, Janko and Altman, Sam and Anadkat, Shyamal and others},
  journal={arXiv preprint arXiv:2303.08774},
  year={2023}
}

@inproceedings{acharya2026towards,
  title={Towards Building a Standard Benchmark for Low-Resource Nepali Information Retrieval},
  author={Acharya, Praveen and Bal, Bal Krishna},
  booktitle={Proceedings of the 49th International ACM SIGIR Conference on Research and Development in Information Retrieval},
  pages={5256--5258},
  year={2026}
}

@inproceedings{bhusal2025enhanced,
  title={Enhanced Retrieval for QA System Tailored for Nepali Legal Documents Focusing on PSC Examinations Using GPT-4 and RAG Framework},
  author={Bhusal, Nabin and Baral, Daya Sagar},
  booktitle={Proceedings of IOE Graduate Conference},
  year={2025}
}

@inproceedings{wagle2026retrieval,
  title={Retrieval Augmented Generation Framework for the Nepali Legal Domain Question Answering},
  author={Wagle, Samir and Adhikari, Abiral and Khanal, Reewaj and Bhandari, Batsal and Manandhar, Prashant and Acharya, Praveen and Bal, Bal Krishna},
  booktitle={2026 8th International Conference on Natural Language Processing (ICNLP)},
  pages={583--592},
  year={2026},
  organization={IEEE}
}

\end{document}